# Automating Nanoindentation: Optimizing Workflows for Precision and Accuracy

Vivek Chawla[1*], Dayakar Penumadu[1], Sergei Kalinin[2]

[1] Department of Civil and Environmental Engineering, University of Tennessee, Knoxville, TN 37996, USA
[2] Department of Materials Science and Engineering, University of Tennessee, Knoxville, TN 37996, USA

## Abstract

Nanoindentation is vital for probing mechanical properties, but traditional grid-based workflows are inefficient for targeting specific microstructural features. We present an automated nanoindentation framework that integrates machine learning, real-time alignment, and adaptive indentation strategies. The system supports three modes: standard automation, feature-based indentation via image-to-coordinate mapping, and large-scale indentation with full x, y, and z-axis alignment. A key challenge—precise sample positioning across imaging and indentation stages—was addressed by correcting travel distance errors (2.5–6 µm) through pixel-to-micron calibration, reducing alignment errors to sub-micron levels. Benchmark tests demonstrate phase-specific and orientation-guided indentation enabled by self-organizing maps and macro imaging. This framework significantly enhances precision, reduces user intervention, and enables efficient, targeted characterization of complex materials. Furthermore, by establishing a direct interface between nanoindentation systems and Python-based automation frameworks, we anticipate these workflows can be adopted across most existing nanoindenter platforms. This work lays the foundation for next-generation autonomous mechanical testing workflows tailored for microstructurally complex materials.

[*] Corresponding author. Tel: +1 315-800-8228 E-mail address: vchawla@vols.utk.edu

Nanoindentation is a key technique for characterizing the mechanical properties of materials, playing a critical role in fields ranging from metallurgy and ceramics to biomaterials and thin-film coatings. Since the seminal work by Oliver and Pharr [1,2], nanoindentation has been instrumental in quantifying hardness, elastic modulus [3,4], and time-dependent mechanical responses such as creep and viscoelasticity [5–9]. The technique has significantly contributed to advancing structural materials, enabling a deeper understanding of deformation mechanisms in metals [6,10–13], ceramics [14–18], polymers [19–21], and composites [22–26]. It is widely used in assessing the mechanical behavior of microelectronics [27–31], biological tissues [32,33], and emerging materials such as high-entropy alloys [34,35] and nanostructured materials [36].

Despite its versatility and broad use across multiple domains, conventional nanoindentation workflows remain constrained by their sampling approaches. Most measurements are conducted using either single-point indentation, which provides localized mechanical characterization, or grid-based mapping, where indentations are arranged in a predefined rectangular pattern. A fundamental limitation of nanoindentation is its irreversible nature—each measurement creates permanent deformation in the material. Consequently, mapping-based approaches often result in significant inefficiencies, particularly when targeting specific microstructural features such as grain boundaries, interfaces, or inclusions. Since only a fraction of the acquired measurements correspond to regions of interest, large portions of the dataset may contribute little to the scientific objective.

To address these inefficiencies, high-throughput nanoindentation systems, such as those developed by FemtoTools and other commercial providers, have emerged, offering improved automation and throughput. However, these advancements do not fundamentally alter the underlying sampling strategy, which remains largely predefined rather than adaptive. A promising avenue for improving nanoindentation efficiency is the adoption of machine learning (ML)-enabled active sampling strategies. Unlike conventional grid-based approaches, ML-driven workflows enable targeted characterization of microstructural elements either by leveraging prior knowledge of relevant features or by dynamically identifying regions of interest based on specific mechanical response criteria. Previously, such workflows have been extensively developed in electron and scanning probe microscopy [37–42] to optimize data acquisition in real time.

However, a key challenge in translating such ML approaches to nanoindentation lies in the inherent complexity of its measurement system. Unlike scanning probe or electron microscopy, where the imaging and spectroscopic measurement processes are inherently integrated via the same probe, nanoindentation relies on an optical system for identifying location of interest and related sample alignment and an independent mechanical probe (spherical, Berkovich, flat indenters) for indentation. This separation necessitates precise control over sample positioning, alignment, and focusing—critical factors that must be addressed for automated experimentation. The successful implementation of ML-driven nanoindentation will therefore require robust engineering controls that enable automated sample point selection, adaptive focusing, and real-time measurement optimization.

Here, we report the development of an ML-enabled nanoindentation framework that integrates automated sample alignment, active learning-based sampling, and precision benchmarking of mechanical property measurements. Our approach leverages commercially available nanoindenters and open-source Python libraries, ensuring broad applicability across different instrument platforms. We demonstrate that incorporating ML-driven decision-making into nanoindentation not only improves measurement efficiency but also enables the systematic exploration of mechanical property variations across complex microstructures. Furthermore, by establishing a direct interface between nanoindentation systems and Python-based automation frameworks, we anticipate that this work will catalyze the development of next-generation autonomous workflows for the mechanical characterization of materials considering combinatorial material science.

# Results
## Nanoindentation Workflow and Error Budget

Nanoindentation experiments typically follow a structured workflow that begins with sample mounting and proceeds through imaging, transition, indentation, and data analysis. This section focuses on the details of the indentation workflow where the sample is first examined under an optical lens before indentation, offering a clear understanding of the associated processes, parameters, and potential errors.

The process begins with securely mounting the sample to ensure stability and flatness during indentation. Next, the sample is examined under the nanoindenter's optical lens. Depending on the experimental objectives, this may involve local optical imaging to target specific features like grain boundaries or interfaces, global macro imaging for large-scale mapping, or moving across multiple locations to perform rectangular grid-based indentations over the entire sample.

Following the imaging step, the sample is transitioned from the optical lens to the indenter for the execution phase. This movement involves a series of carefully controlled steps to minimize errors and ensure safety. Initially, the Z stage ascends, and the XY stage moves the sample from the optical lens to a position beneath the indenter. The indenter is positioned not at the intended test point but at a nearby user-defined location. This location is deliberately chosen to prevent unintended indentation while identifying the surface. Once positioned, the Z stage lowers to locate the surface of the sample. At this point, the Z stage remains stationary, but the actuator tip retracts slightly to create space. The XY stage then moves to the indentation location, and the actuator tip descends to perform the indentation.

From a definitional perspective, it is important to distinguish between different Z positions. The "extension" refers to the position of the Z stage, which is typically motor-controlled and capable of significant movement. In contrast, "displacement" describes the position of the actuator tip, which has a limited range of motion, usually less than 80 µm. During the indentation phase, the system collects data at the selected points, measuring parameters such as load, depth, stiffness, and hardness. For cases involving small-scale features, such as grain boundaries, the destructive nature of nanoindentation necessitates careful planning to minimize damage to critical areas. The data collected from the indentation process includes force-displacement curves, hardness, modulus, and Z position.

During the workflow, several sources of error can affect measurement accuracy, requiring careful quantification. The first major source is the initial travel distance error, which arises when the sample moves from beneath the optical lens to the indenter. In our case, this movement involves approximately 43,000 µm in the X direction and 50 µm in the Y direction, referred to as the initial travel distance in X and Y, respectively. Since this is the largest displacement in the process, it introduces the most significant systematic offset, affecting all subsequent measurements. After this initial offset, indentations are performed in a grid pattern, where indentation positional errors occur due to discrepancies between the intended and actual spacing between indents. If the ideal spacing is defined as dx and dy, but the actual movements are $dx_1$ and $dy_1$, the positional error in the X and Y directions is given by $|dx_1 - dx|$ and $|dy_1 - dy|$, respectively, as illustrated in Figure 1. Finally, height-based positional errors introduce additional variability, as deviations in contact height at a fixed XY position, even by a small margin, can result in further misalignment and contribute to errors in both the X and Y directions.

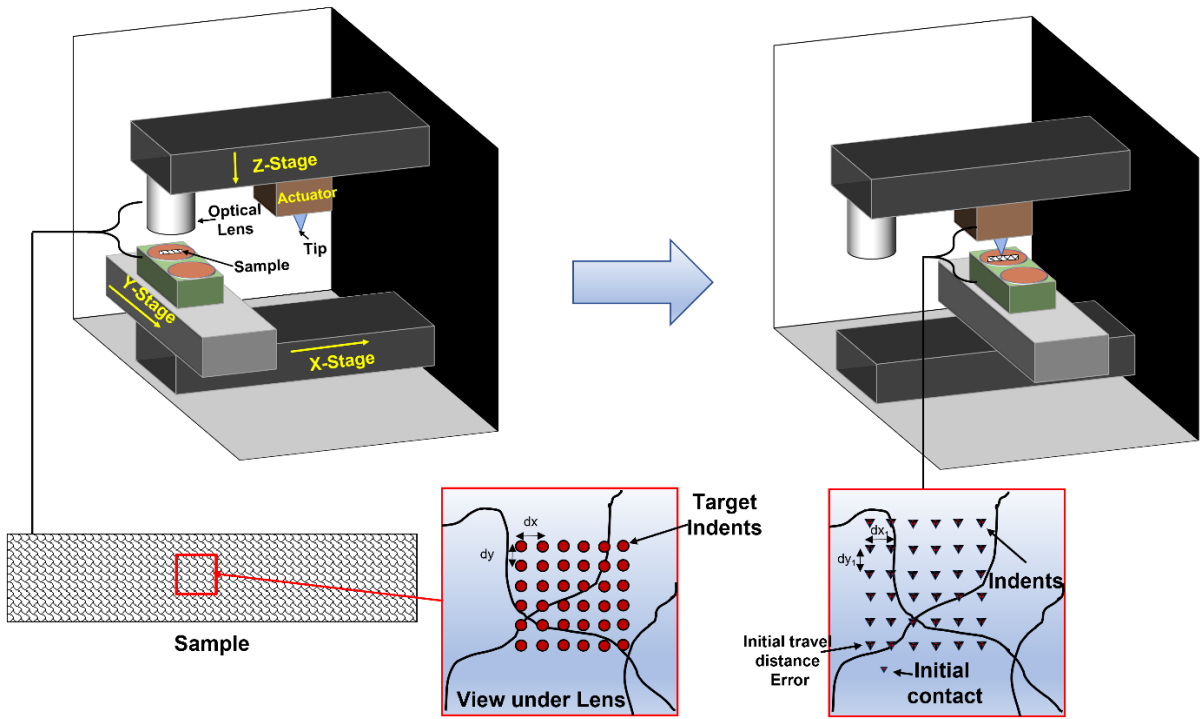

*Figure 1: Schematic of a typical nanoindenter workflow: The sample is initially examined under an optical lens to identify the indentation locations. It is then positioned under the indenter, where contact is established using an offset point before performing the indentations.*

## Positional Error and Scale Characterization

A comprehensive characterization of the nanoindenter system is presented in the Supplementary Information. This includes assessments of initial travel distance errors, positional accuracy, XY scale calibration, and image distortion. We find that the first indent after a motion cycle exhibits a maximum error of 2.5–6 µm, while relative errors between subsequent indents are below 1 µm. Calibration of the optical system revealed anisotropic image distortion, with pixel-to-micron conversion factors of 5.581 pixels/µm along the X axis and 5.891 pixels/µm along the Y axis. Z-positioning errors due to sample tilt and tip misalignment were found to induce systematic XY errors, which scale linearly with height variation. Motion accuracy was further evaluated through distance-dependent tests and showed that positional errors increase with travel distance, reaching up to 10 µm over large displacements. For small displacements (<25 µm), errors remained below 500 nm. Finally, XY hysteresis was found to be minimal (~300 nm), likely due to active backlash compensation. These findings highlight key sources of error relevant to accurate spatial localization.

# Building an Automated Nanoindenter

## Gaining Control Over the Indenter Interface

To build an automated indenter, direct back-end access to the indenter controls is ideal. However, this is not available in most nanoindenter setup. Instead, we circumvent this limitation by leveraging front-end automation tools that allow us to control the indenter through the user interface, just as a human operator would. This approach involves using PyAutoGUI, a Python library for programmatically controlling keyboard and mouse inputs. Figure 2a presents the iMicro nanoindenter's screen interface, where we have access to key parameters, including the x, y, and z coordinates and their corresponding movement controls. Through this interface, indentation at a precise location can be achieved. One key advantage of this method is that the center of the screen acts as the origin, allowing for a coordinate system where indentations can be programmed relative to this reference point. In addition to controlling the indenter, it is crucial to read the current position coordinates. This is accomplished through OCR (Optical Character Recognition) libraries, which extract text data from the screen, allowing us to track position updates dynamically.

## Implementing Automated Movements in X and Y Directions

An automated indenter must be capable of controlled movement along the x and y axes. Figures 2b demonstrate how automated relative movement is implemented. For instance, we can programmatically shift the indenter 50 µm to the left by issuing the corresponding keyboard or mouse inputs through PyAutoGUI. Using this technique, we can perform systematic indentations at predefined locations, ensuring consistency in data collection and eliminating operator variability.

## Feature-Based Indentation Using Image Processing

Beyond coordinate-based indentation, we also leverage the indenter's optical imaging capabilities to enable feature-based indentation. The indenter provides real-time images of the sample surface, which allows us to select indentation locations based on specific microstructural features. This is achieved through screen capture techniques followed by image processing using libraries such as scikit-image or OpenCV. A basic approach involves identifying contrast variations or texture differences to locate specific regions of interest. More advanced implementations could incorporate deep learning models, such as Deep Kernel Learning (DKL), as demonstrated in literature [37], to enable autonomous decision-making in indentation site selection. Once a feature of interest is identified, we determine its pixel coordinates within the captured image. These pixel positions are then converted into micron-scale coordinates using the pixel-to-micron conversion established earlier. This is done relative to the indenter's origin (crosshair). Figures 2c illustrates this process, where an indentation point is selected based on microstructural features visible in the optical view.

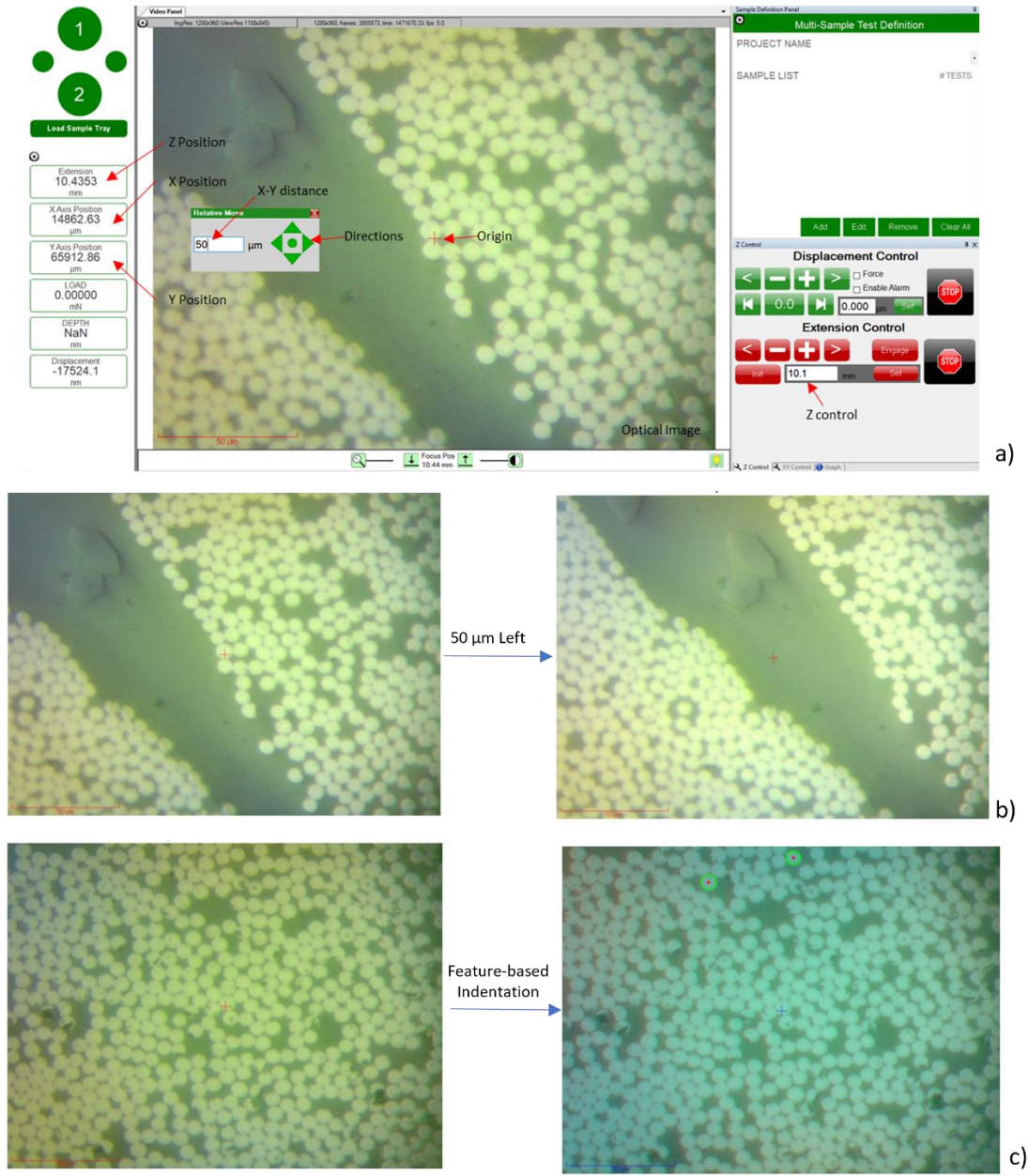

*Figure 2: User controls and automation in the nanoindenter using the PyAutoGUI library. (a) Probes and user controls available for the nanoindenter. (b) Demonstration of movement in any x-y direction by simulating keyboard input. (c) Access to features visible through the nanoindenter's optical lens and corresponding feature-based indentations.*

**Aligning Macro and Micro Images for Enhanced Localization**

A unique workflow in nanoindentation experiments is aligning the macro view (e.g., an optical microscope or SEM image) with the micro view obtained from the nanoindenter's optics. This alignment ensures accurate positioning for multi-scale material analysis, which is typically infeasible without automation. The process involves establishing a transformation between the two imaging systems using reference points on the sample. A detailed description of the alignment procedure, including XY and Z alignment steps, is provided in the supplemental information. After alignment, an automated indentation process is demonstrated in Figures 3. Figure 3a shows four preselected locations, while Figure 3b presents a scenario where the indenter moves to one of these locations but does not perform z-alignment, resulting in a blurry image. Finally, Figure 3c demonstrates the result after z-alignment, where the image comes into focus. This approach ensures that indentation locations chosen in the macro-scale image can be accurately transferred to the nanoindenter's coordinate system, enabling multi-scale correlative analysis.

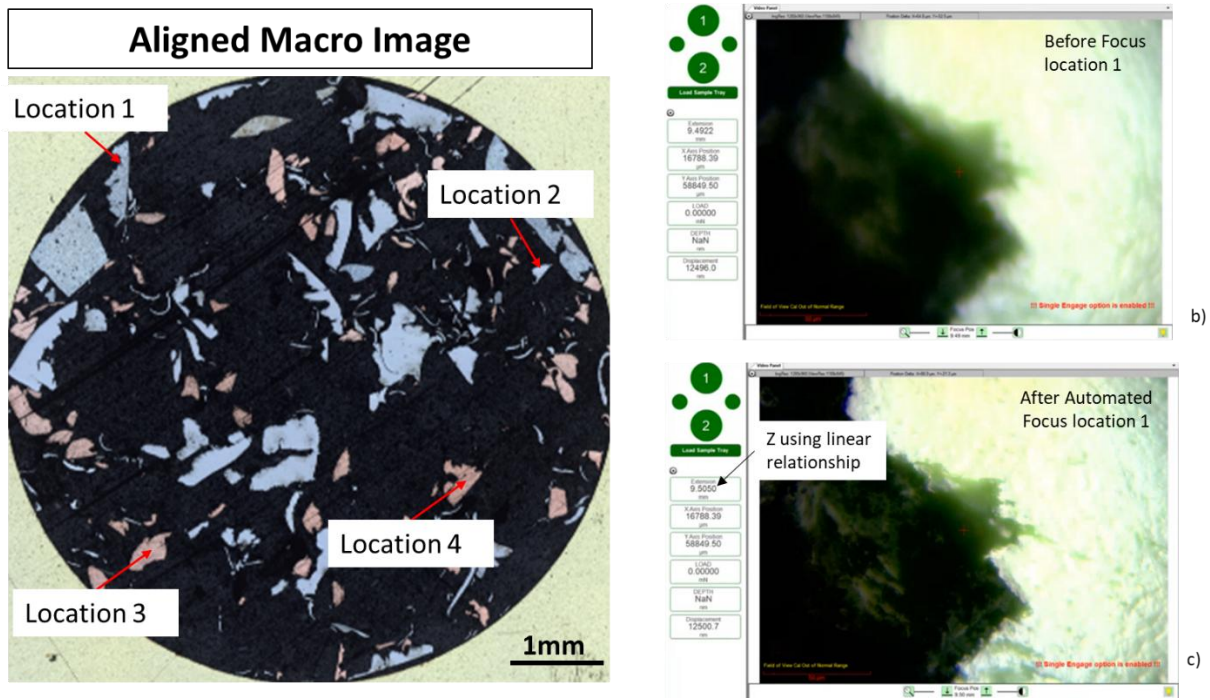

*Figure 3: (a) Aligned macro image with randomly chosen four target locations. (b, c) Example of the indenter automatically moving to one of the target locations—(b) before focus adjustment and (c) after achieving focus.*

# Discussion

## Benchmark Problem: Feature-Based Segmentation and Automated Indentation

To demonstrate the effectiveness of feature-based segmentation for automated indentation, we present a benchmark problem using a stone sample. While we do not explicitly identify the material, we showcase its composition using optical imaging and energy-dispersive spectroscopy

(EDS) mapping. Figure 4 presents the optical image of the region of interest (ROI) along with its corresponding EDS elemental maps, revealing the presence of sodium (Na), potassium (K), barium (Ba), silicon (Si), iron (Fe), magnesium (Mg), aluminum (Al), and calcium (Ca). The sample consists of four distinct phases. To identify these phases, we employ color thresholding to distinguish the different material regions based on variations in elemental composition. This problem serves as an excellent starting point for demonstrating feature-based segmentation, allowing us to categorize regions of interest before conducting targeted nanoindentation.

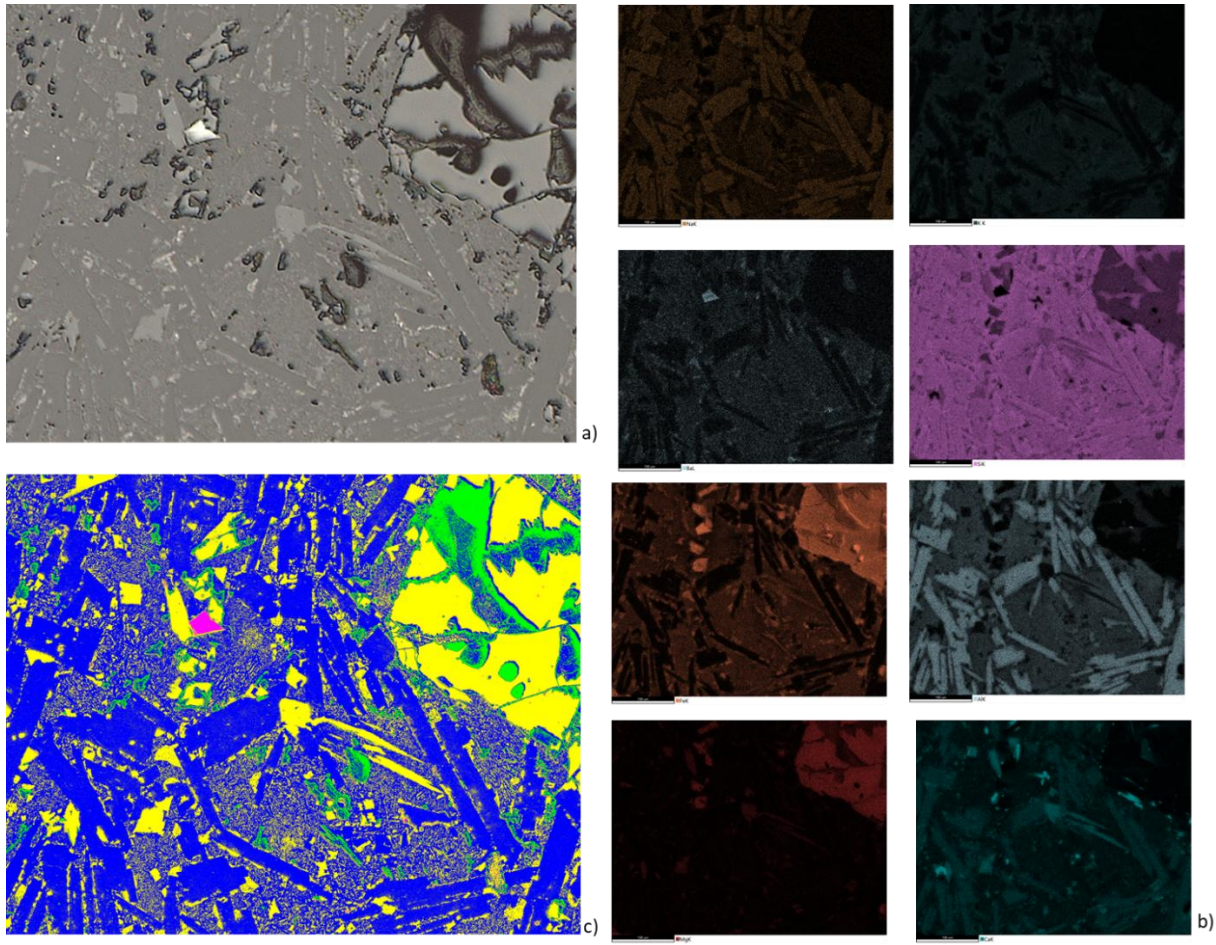

*Figure 4: Phase identification using EDS and optical microscopy (a) Optical microscope image of a stone with a field of view similar to that of the nanoindenter. (b) EDS scans showing the distribution of different elements. (c) Segmentation of the optical image using color-based thresholding to identify different phases.*

After segmenting the image into distinct phases, we proceed to perform automated nanoindentation using the Self-Organizing Feature Map (SOFM) algorithm. SOFM is a type of unsupervised neural network used for clustering and visualization. It maps high-dimensional data into lower-dimensional space while preserving topological properties. Unlike conventional clustering methods such as k-means, SOFM maintains the spatial structure of data, making it highly suitable for optimizing indentation locations while minimizing travel distance. In Figure 5, we apply SOFM to the thresholded image, generating a map of optimal indentation locations. The key steps involved are:

1. Input Representation: The segmented image is used as an input, where each pixel corresponds to a potential indentation site.
2. Cluster Formation: SOFM organizes the indentation sites into clusters, ensuring that each phase is represented while minimizing redundant travel distance.
3. Indentation Planning: The closest indentation points from each phase are selected to establish an optimized path for indentation.

Using this approach, we transition from manual selection of indentation sites to a systematic, image-based method, improving efficiency and consistency. Once indentation locations are selected in the image, their pixel coordinates are converted into real-world XY positions relative to the nanoindenter's coordinate system. This transformation is crucial, as it allows for direct execution of indentation experiments based on image analysis rather than relying on manual input.

Following indentation, we conduct post-indentation microscopy to validate the accuracy of our targeting. The images confirm that the system successfully placed indents at the intended locations, verifying the reliability of the automated indentation process. At first glance, this process may seem straightforward. However, it marks a significant departure from traditional nanoindentation workflows. Conventionally, nanoindentation is often performed using a rectangular grid-based analysis, where indentations are systematically placed across the sample to analyze different phases and their distribution. While this method ensures broad coverage, it is data and time-intensive and may not adequately capture complex solution spaces where critical features fall between grid points. An alternative approach involves manually selecting features of interest, determining their positions relative to the origin, and inputting indentation locations. However, this manual approach is time-consuming and impractical for samples with numerous scattered features. Automating this process allows the indenter to intelligently detect and indent features of interest, significantly enhancing efficiency and precision. Our study showcases a paradigm shift where feature-based indentation can be automated at scale, removing the need for labor-intensive manual selection. This lays the groundwork for further advancements in automated decision-making for nanoindentation. The only remaining challenge is how the nanoindenter itself can autonomously make decisions about indentation locations based on material properties and experimental objectives.

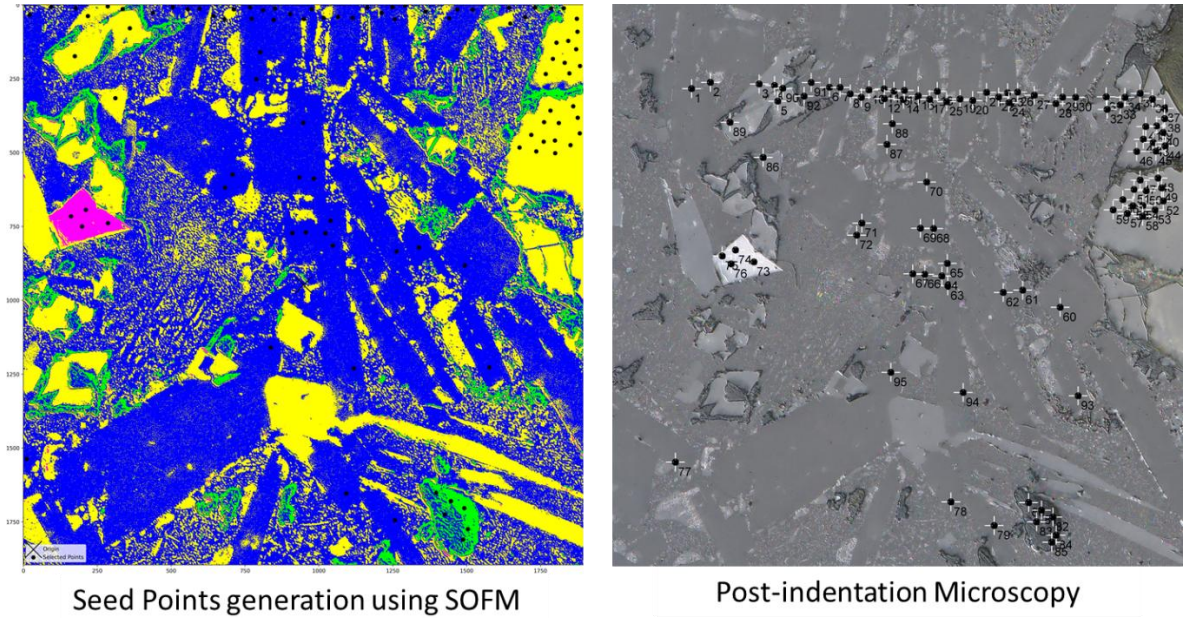

*Figure 5: Benchmark example of automated indentation positioning. (a) Ideal indentation locations identified using the Self-Organizing Feature Map (SOFM). (b) Post-indentation optical microscopy image confirming the indentations.*

## Benchmark Problem: Feature Based indentation at macro scale

Characterizing local mechanical properties of carbon fibers in randomly oriented carbon fiber-reinforced composites is a particularly challenging problem—not only because carbon fibers are orthotropic, but also due to the inherent variability in properties for the same fiber orientation due to the use of recycled fibers in many composites, which often contain a mix of carbon fibers derived from different precursors with varying microstructures. As a result, understanding the distribution of mechanical properties as a function of fiber orientation is a complex and non-trivial task. While computational methods such as finite element modeling (FEM) and homogenization are commonly employed to estimate the effective modulus of anisotropic carbon fibers, these methods lack the direct experimental data necessary for accurate modeling and validation. Experimental approaches like nanoindentation offer a unique solution by providing localized mechanical property data; however, their implementation in these systems is highly complex due to the randomness of fiber orientation and the need for statistically significant data from numerous fibers. Therefore, in the next problem, we demonstrate how the automated feature-based nanoindentation approach that leverages macro-scale optical imaging to guide the nanoindenter can be used to solve such a problem. A macro image of the composite sample, shown in Figure 6a, reveals randomly distributed carbon fibers surrounded by matrix material and pores. Image segmentation techniques are applied to this macro image to identify and isolate individual fibers based on their grayscale intensity and morphological features. Simple thresholding is employed for initial segmentation, followed by filtering criteria based on fiber size to ensure that only fibers with desired orientations are selected for indentation.

For this demonstration, we focus on three distinct fiber types categorized by their cross-sectional area: 30–40 µm² for fibers oriented at 0°, greater than 1000 µm² for fibers at 90°, and 50–60 µm² for fibers with an intermediate orientation between 0° and 90°. These orientations correspond to the longitudinal, transverse, and intermediate directions, respectively. The primary objective of this study is to demonstrate how an automated indenter can assist in addressing this classification problem. Rather than focusing on precise fiber orientation identification, which has been extensively studied in the literature [43,44], we aim to explore how an automated indenter can facilitate solving previously unfeasible problems, given that the user has access to orientation distribution. Figure 6b highlights the selected fibers in red, green, and blue, corresponding to the three orientations. While a full characterization would involve a broader range of fiber orientations and multiple indentations across numerous fibers, the current example aims to showcase the workflow and the potential of the automated system. Once the fibers are identified, the automated nanoindenter is programmed to perform indentation at a single location within each fiber. The NanoBlitz mode is used for this purpose, which performs high-speed grid indentation across a selected region of the fiber. Unlike conventional nanoindentation that provides detailed load-depth curves for each indentation, NanoBlitz generates modulus and hardness maps, offering a rapid overview of the spatial variation in properties. The indenter view for each of these fibers and the corresponding modulus maps are shown in Figure 6c-h. The results, reveal distinct differences in the elastic modulus for each fiber orientation. The modulus is highest for fibers oriented along the 0° direction, reflecting the high stiffness of carbon fibers in their longitudinal direction. For the intermediate orientation fibers, the modulus is lower but still higher than that of the 90° fibers, indicating the influence of both axial and transverse properties. The 90° fibers exhibit the lowest modulus, as expected, given their transverse orientation, where the matrix material predominantly governs the mechanical response.

This automated workflow does not merely facilitate the collection of orientation-specific data; it transforms the feasibility of solving an otherwise unsolvable problem. With minimal user intervention, the nanoindenter is capable of autonomously identifying fiber features and performing targeted indentations, generating high-density datasets that can be analyzed for orientation-property relationships. The example described here serves to illustrate the capabilities of automation. But we can go a step further. By integrating a Bayesian decision framework, the mapping of mechanical properties as a function of orientation can be optimized and refined iteratively. Moreover, the role of smaller factors of variations such as residual stresses could also be disentangled. Such a framework would enable data-driven decision-making, identifying which fiber orientations need further characterization and reducing uncertainty in the orientation-property map. In practical terms, this means the nanoindenter could iteratively refine the orientation-property map, making real-time decisions on where to indent next to maximize the information gained. This would transform the problem from a purely experimental challenge to a smart, adaptive process driven by both experimental data and probabilistic modeling. However, that is beyond the scope of this study and a topic for future work.

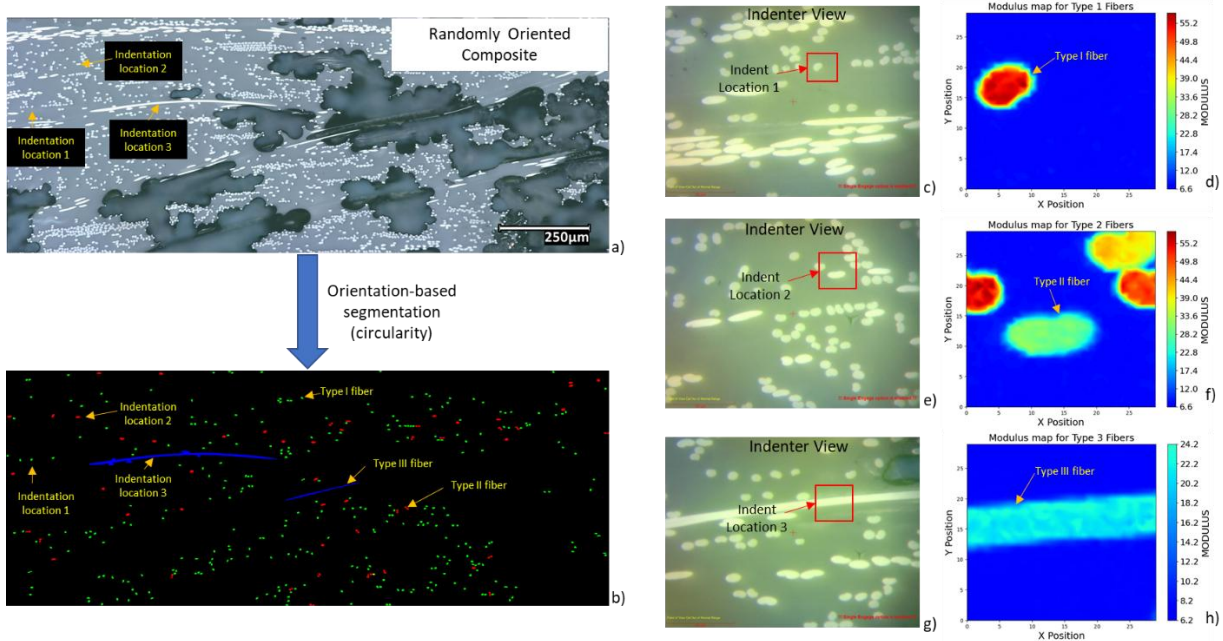

*Figure 6: Benchmark example II: Indentation on randomly oriented carbon fibers. (a) Macro image showing randomly oriented carbon fibers within a matrix containing voids. (b) Classification of fibers into three types based on size, with color-coded distributions and corresponding target indentation locations. (c, d) Nanoindenter optical image and resulting NanoBlitz contour plot for a Type I fiber. (e, f) Nanoindenter optical image and contour plot for a Type II fiber. (g, h) Nanoindenter optical image and contour plot for a Type III fiber.*

## Improving the X-Y stage performance

For most instruments, performance improvement is primarily achieved by manufacturers, and enhancing performance from the user's end may seem limited in impact. However, since users are constrained by the instruments they have, such efforts are not necessarily without value. High-precision indentation applications, such as pillar-splitting and push-out testing, demand sub-micron stage resolution, often pushing the limits of stepper motor-controlled X-Y stages. One potential method to improve X-Y stage accuracy during nanoindentation is to adopt an incremental approach when transitioning between the optical lens and the indenter. Rather than making a direct move between the two, which can introduce significant errors, incremental steps which have a smaller error associated with them can reduce error accumulation at each stage. For instance, if the total distance between the optical lens and the indenter in the x-direction is 43,000 µm and 63 µm in the y-direction, we could move in 1000- µm steps. Assuming a maximum positioning error of approximately 2.5 µm per step, this would require 43 moves in the x-direction and one in the y-direction. Each movement, including backlash removal, takes about 20 seconds, resulting in a total movement time of around 15 minutes—an acceptable compromise if higher accuracy is achieved. However, this incremental approach has a significant drawback. Positioning errors typically follow a Gaussian distribution with a standard deviation (σ) for each step. With N-incremental steps, the overall standard deviation of the error of the trace becomes $\sigma * \sqrt{N}$ . As a result, the cumulative error grows larger with each step, undermining the accuracy improvement and negating the primary purpose of the incremental approach. Moreover, the additional time required makes this method inefficient for high-throughput experiments.

A more robust strategy for improving accuracy in critical cases involves leveraging nearby features as reference points. If no suitable feature exists, we can create one by making a small indent near the desired location (Figure 7a). A NanoBlitz grid can then be performed over this feature to generate a high-resolution contour map based on relevant data—such as modulus, hardness, or Z-position. This map provides a reliable reference for aligning the feature with the target location. Alignment can then be performed manually or, ideally, through an automated process. Once the alignment is complete, the exact position beneath the indenter can be determined, ensuring that we are within few tens of microns of the feature of interest. Once the reference position is established, we can precisely move to the desired indentation location. Although the above process can be done manually, repeating the process multiple times for different locations is tedious and prone to human error. Moreover, this requires blindly using the indenter while the sample is directly underneath it. Therefore, a small human operating error can result in actuator damage. Automating this process significantly reduces the effort required along with ensuring the safety of the equipment while improving the stage resolution. While incorporating a NanoBlitz grid (~350 seconds) and additional movements adds time to the workflow, the gain in accuracy and reliability outweighs the extra time investment for cases where positional accuracy on indentation is of prime importance.

Figure 7a shows a nanoindenter image under the optical lens, with an indent used as a reference feature located 25 µm to the left of the intended indentation point. Figures 7b illustrates the NanoBlitz grid (400 indentations over 20 µm * 20 µm) and the subsequent alignment process. Once aligned, the known x- and y-scale factors of the indenter allow for precise movement to the target location, minimizing positional errors and ensuring consistent results. Automation plays a crucial role in streamlining this workflow. The proposed workflow significantly improves positional accuracy, reducing errors to less than 1 µm, making it particularly suitable for high-precision applications such as indentation near phase boundaries, pillar splitting, and push-out testing.

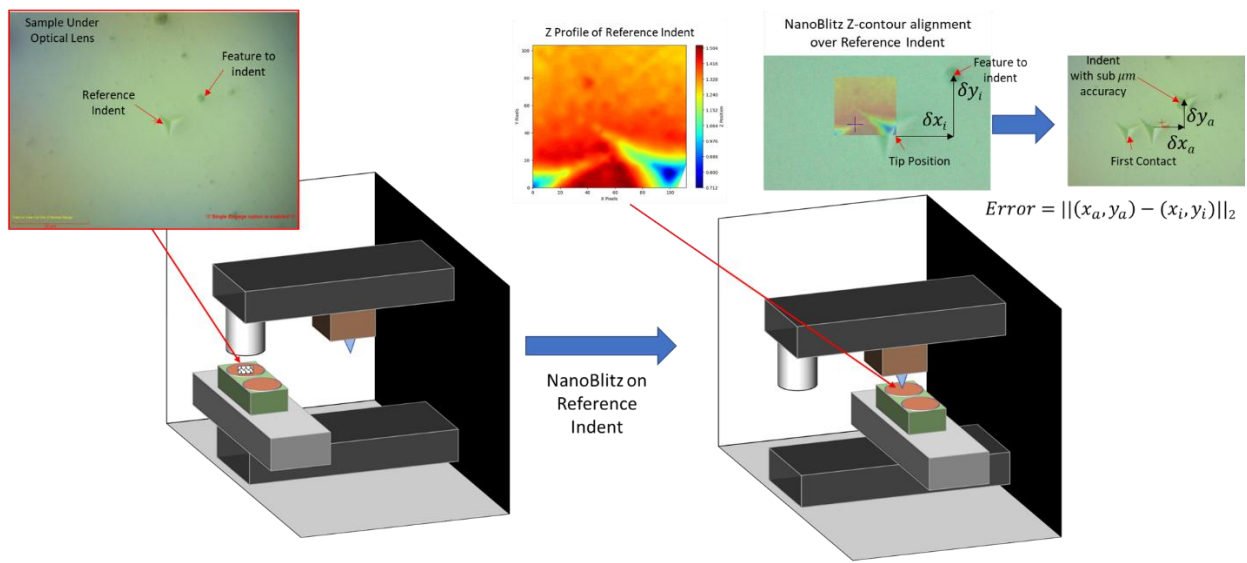

*Figure 7: Automated Workflow for High-Accuracy Indentation. (a) Sample under the optical lens, showing both a feature to be indented and a nearby reference indent. (b) Workflow demonstration: the sample moves beneath the indenter, and NanoBlitz is performed on the reference indent. Using the Z-contour plot from NanoBlitz and the original optical image, the tip's position*

*relative to the sample is identified, mitigating initial travel distance errors. The system then moves to the target feature for indentation.*

In summary, we introduce an automated nanoindenter framework that transforms indentation experiments by offering three distinct modes: (1) traditional workflow automation, (2) feature-based indentation via image recognition and pixel-to-micron conversion, and (3) large-scale, macro-image–guided indentation with automated XYZ alignment. Each mode overcomes specific experimental bottlenecks. Despite hardware limitations, our automated alignment strategy successfully reduces errors to sub-micron levels in high-accuracy applications. We demonstrate the framework's versatility through two benchmarks: phase-specific indentation using self-organizing feature maps, and orientation-dependent fiber indentation mapping mechanical property variations. This platform bridges manual operation and full automation, empowering researchers to design complex, high-throughput studies across scales. With further hardware refinements, it promises even higher resolution and broader applicability, paving the way for more efficient and reproducible materials research.

## Methods

The nanoindentation experiments were conducted using an iMicro nanoindenter from KLA, which was mounted on a vibration isolation table and enclosed within an environmental chamber to minimize external noise. A Berkovich indenter tip, sourced from Synton MDP, with a cone angle of 65.3°, was used for all measurements. Two distinct materials were examined for benchmark problems in this study. The first material is a stone sample being evaluated as a potential source rock for basalt fiber production. The second material is a randomly oriented carbon fiber composite, fabricated using recycled carded fibers in a polyphenylene sulfide (PPS) matrix[45].

## Author contributions

V.C. formulated the automation code, demonstrated its performance on two benchmark problems, and drafted the original manuscript; D.P. conceptualized the underlying idea and guided its revision and refinement; and S.K. provided the original conceptualization of the project, offered strategic guidance throughout, and—together with D.P.—revised and finalized the manuscript.

## Acknowledgements

V.C. would like to acknowledge Dr. Stephen Puplampu for his assistance with the experimental setup. This research was primarily supported by the National Science Foundation Materials Research Science and Engineering Center (MRSEC) program through the UT Knoxville Center for Advanced Materials and Manufacturing (DMR-2309083). EDS was performed at the Institute for Advanced Materials & Manufacturing (IAMM) microscopy Facility, located at the University of Tennessee, Knoxville.

## Competing Interests
The authors declare no competing interests.

# Supplementary Material

# Performance Evaluation of the Indenter

To ensure the reliability and precision of nanoindentation experiments, we perform a comprehensive series of evaluations to characterize the performance of the indenter. These assessments encompass multiple testing methodologies, each designed to quantify specific sources of error and evaluate the system's accuracy and repeatability. A critical aspect of this characterization involves analyzing the motion of the indenter along the x, y, and z axes, as well as identifying image-based errors that could influence measurement fidelity. By systematically examining these factors, we aim to establish a robust error profile for the nanoindentation system.

## Initial Travel Distance error and positional error

All distance measurements were performed using a Keyence VHX-7000 optical microscope, which provides high-resolution imaging. To minimize systematic errors in later analyses, we first conducted an alignment procedure (Figure S1a, S1b). This involved placing reference or landmark indents on the sample surface and imaging them under the microscope. Horizontal and vertical guide lines were used to precisely position the indents, ensuring accurate alignment prior to any quantitative measurement.

Before evaluating positional and initial travel distance errors, we performed XY stage calibration as specified by the manufacturer. This process involved selecting a reference point, executing an indentation, and measuring the deviation between the expected and actual indentation positions. The system was then adjusted to compensate for this discrepancy, improving accuracy for subsequent motions.

Following calibration, an indentation pattern centered on the reference point (fiber center in Figure S1c) was executed. This pattern consisted of indents arranged either in a star-shaped configuration or along two concentric circles with radii of 20 µm and 30 µm. The actual indent positions were measured to evaluate deviations in motion and positioning accuracy.

Despite calibration, the first indentation after initiating a new motion cycle exhibited measurable travel distance errors. As shown in Figure S1c and S1d, the actual positions were compared against expected locations using the calibrated pixel-to-micron scale. These measurements were repeated across multiple samples to ensure consistency. Results showed that maximum initial travel distance errors ranged from 2.5 to 6 µm, while subsequent indentations exhibited much smaller relative positional errors—typically within a few hundred nanometers.

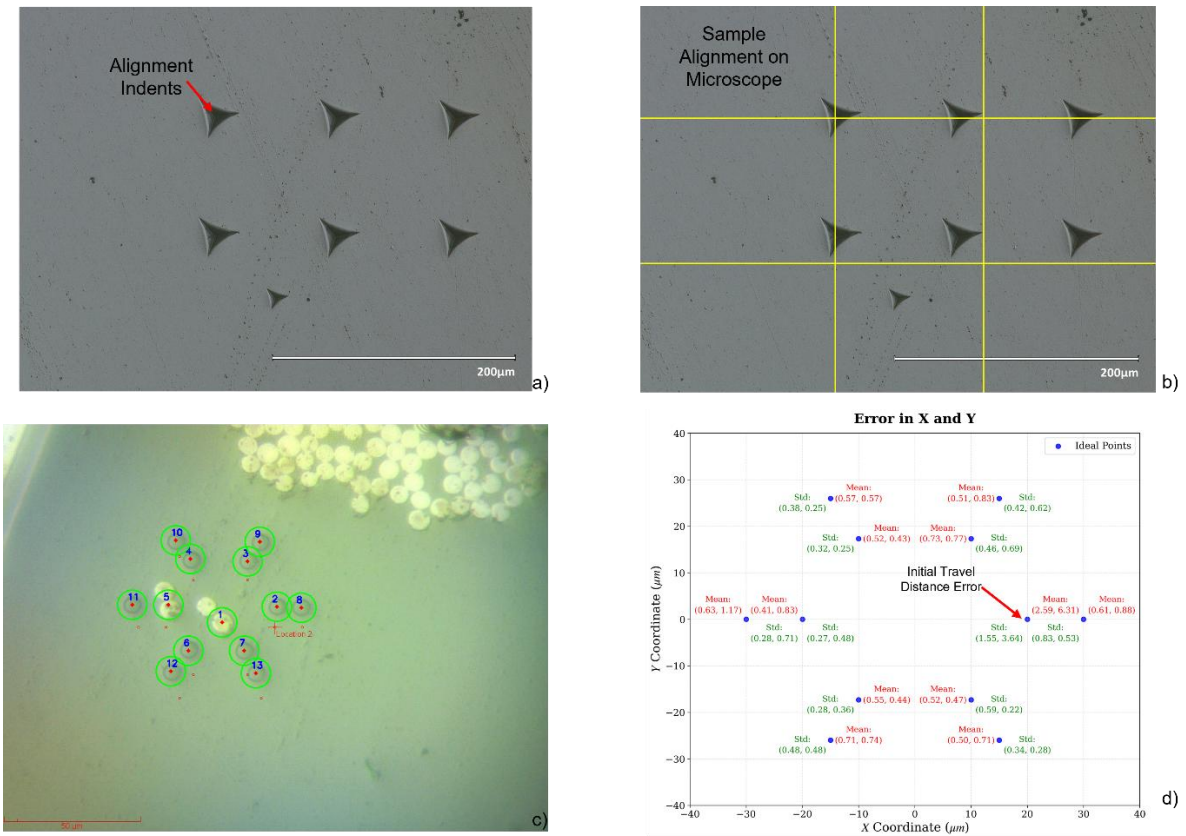

*Figure S1: Calibration and performance of the iMicro nanoindenter. (a, b) Sample alignment using reference indents observed under a high-magnification optical microscope. (c) Indentation pattern after the experiment. (d) Error analysis showing the maximum error associated with the initial travel distance.*

## XY Scale Calibration and Image Distortion

To calibrate the XY scale and assess image distortion, indentations were placed at predefined locations to generate visible markers on the nanoindenter's imaging screen (Figure S2a). The distances between these markers were measured in pixel units along the X and Y directions using the nanoindenter's optical view. These pixel-based measurements were then calibrated against true distances obtained from high-resolution images captured using the Keyence VHX-7000 microscope (Figure S2b), which provided micron-scale accuracy.

This calibration process yielded pixel-to-micron conversion factors of 5.581 pixels/µm in the X direction and 5.891 pixels/µm in the Y direction (Figures S2c and S2d). The discrepancy between these values confirmed the presence of anisotropic image distortion in the optical system. To ensure consistency and robustness, this procedure was repeated at multiple sample locations. All samples were aligned using the previously described protocol.

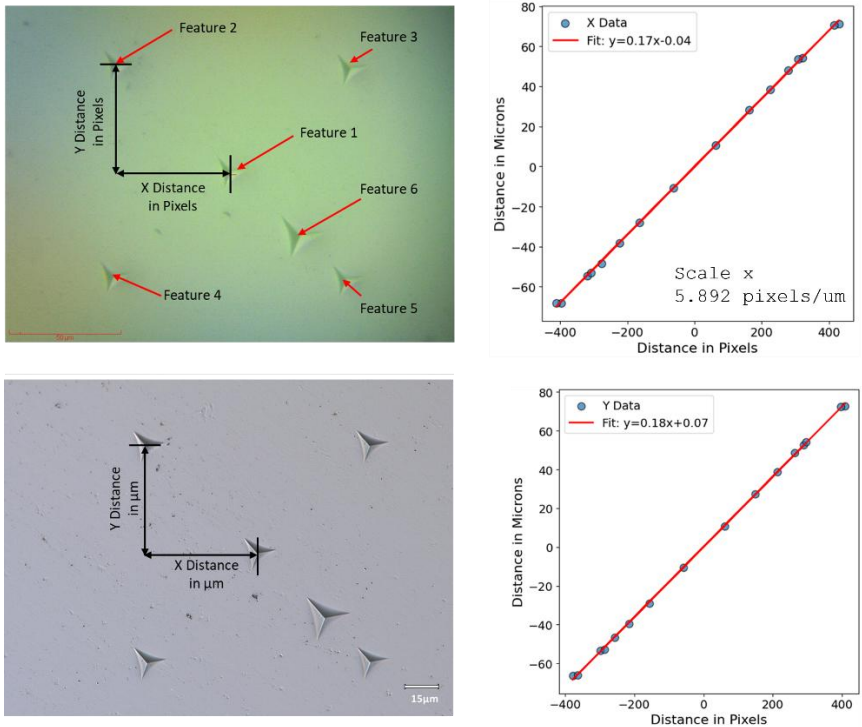

*Figure S2: a) Measurement of distances in pixels using the nanoindenter's optical lens. (b) Measurement of distances in µm using a high-magnification optical microscope. (c, d) Calibration results showing the x- and y-axis scales in pixels per µm.*

## Z- Position and Tilt-Induced Errors

In an ideal setup, sample mounting should ensure a uniform Z position, minimizing height variations across the surface. However, due to sample tilt and surface topography, even slight deviations over large travel distances can introduce inconsistencies in Z positioning. Additionally, no indentation tip is perfectly perpendicular to the sample surface. As a result, if the stage is calibrated to a specific Z height (e.g., 10 mm), but the actual contact occurs at a different height (e.g., 10.5 mm), unintended X and Y positioning errors can arise. To minimize such inaccuracies, the influence of sample tilt and tip alignment must be systematically characterized.

To quantify this effect, we performed indentations at a known feature while varying the Z height across different sample placements. Z height was determined using two methods: optical focus and the contact-based Z value from the indenter. For this analysis, we used the focus-based height, as it is known prior to indentation. A shallow indent was then placed, and its actual position was compared with the expected target. For instance, if the target is 30 µm to the right and below the reference point, but the actual indent appears 28 µm to the right and 29 µm below, this corresponds to an initial travel distance error of 2 µm in X and 1 µm in Y.

Since this misalignment follows a predictable trend, we modeled the relationship between Z height variation and lateral XY error using linear regression (Figure S3). As expected, the relationship is approximately linear, reflecting the proportional shift induced by tip tilt. To further evaluate the residual error distribution, we plotted kernel density estimations (KDEs) for X and Y errors. Interestingly, while the X error follows a unimodal distribution, the Y error is multimodal, indicating possible additional contributing factors. However, since our objective is to quantify rather than explain these deviations, further investigation into their origin is beyond the scope of this study.

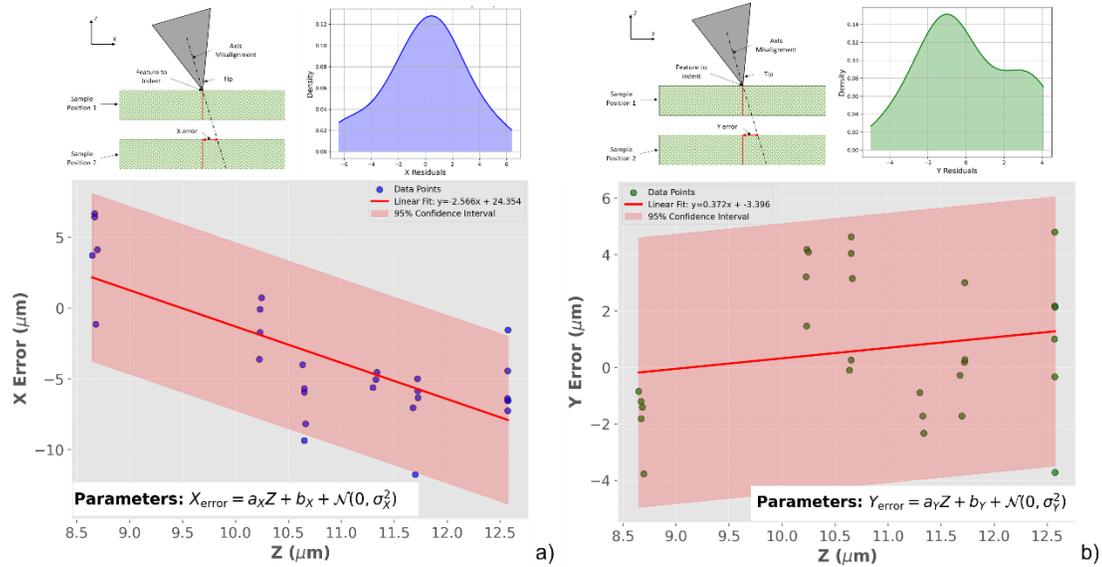

*Figure S2: Error associated with Z-positioning of the sample. (a) X-axis error vs. Z-position with linear fit and 95% confidence interval. (b) Y-axis error vs. Z-position with linear fit and 95% confidence interval.*

## XY Motion Error Characterization

In a PID-controlled system, XY motion errors can vary with travel distance. To characterize this behavior, we performed indentations at randomly selected locations, with target positions sampled from normal distributions centered at zero and standard deviations of 25 µm, 50 µm, 500 µm, and 1000 µm. The actual indentation positions were compared against their prescribed locations to quantify positional errors. In cases where indentations were not detected, placeholder points were inserted but excluded from subsequent distance and error analyses.

The results reveal a clear relationship between travel distance and positioning error. As shown in Figure S4b, optical imaging confirms accurate indentation placement for the 25 µm case. Error scaling is further illustrated in Figures S4c and S4d, which present log-scale plots of error versus travel distance. These indicate that positional errors remain minimal for displacements below 50 µm, but increase significantly at longer distances—reaching up to 10 µm for the largest travel range.

A more detailed analysis shows that for displacements under 200 µm, XY errors remain within 2.5 µm, while for distances below 25 µm, errors are constrained to under 500 nm. These findings highlight the importance of recalibration when operating over large travel ranges, as systematic error accumulation becomes more pronounced. By decomposing the error into linear and nonlinear components, we gain further insight into how travel distance affects positioning accuracy.

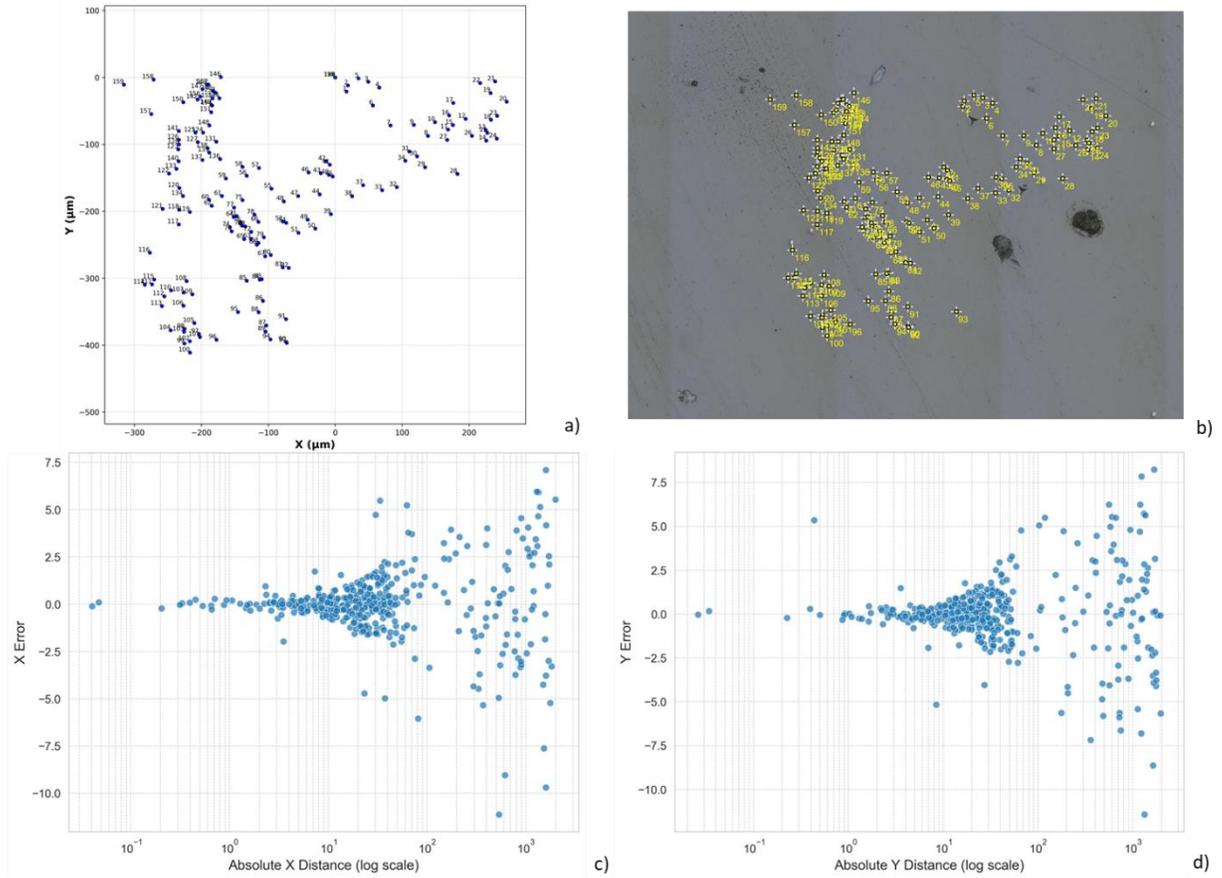

*Figure S3: XY stage error evaluated through random indentation tests. (a) Ideal indentation locations. (b) Post-indentation imaging analysis. (c, d) X-axis and Y-axis errors, respectively, plotted as a function of the distance traveled from the previous indent.*

## XY Hysteresis Error

To evaluate hysteresis error in the nanoindenter's XY motion, we performed controlled movements along an equilateral triangular path, sequentially moving from one vertex to the next and returning to the starting point. The deviation between the initial and final positions quantifies the hysteresis error. Unlike previous assessments, direct indentations were avoided to prevent irreversible surface modification. Instead, we used optical imaging to track positional accuracy.

As shown in Figure S5a, three example triangular paths traced by the system are visualized. Figure S5b displays optical images of the triangle vertices captured by the nanoindenter's lens. The results reveal a small but measurable deviation from the original starting point, confirming the presence of hysteresis. Further analysis shows that the mean displacement error is approximately 300 nm. This low hysteresis error is likely due to active compensation for mechanical backlash after each movement, which effectively reduces cumulative positioning inaccuracies.

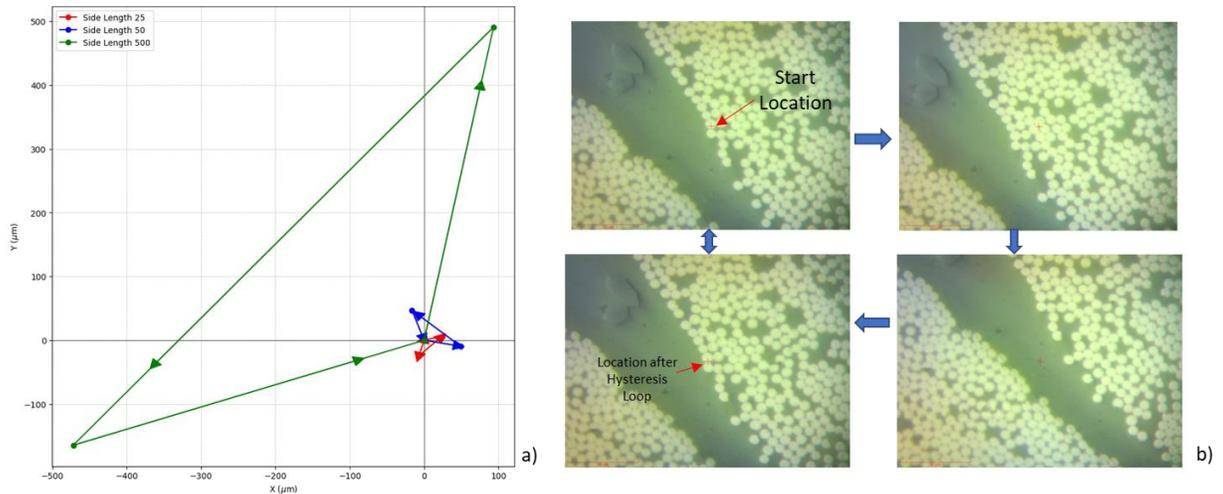

*Figure S4: Hysteresis error evaluated by random movement across an equilateral triangle. (a) Three example paths with different side lengths of the equilateral triangle. (b) Movement from one vertex to another and back to the initial point, visualized using the nanoindenter's optical lens.*

## Aligning Macro and Micro Images for Enhanced Localization
### XY Alignment

To align the x and y coordinates between the macro image and the nanoindenter's optical view, we follow these steps:

1. Select two reference points visible in both the macro image and the indenter's optical view.
2. Measure their coordinates in both images.
3. Calculate the orientation angles (θ for the macro image and φ for the nanoindenter) using the arctangent of the coordinate differences.
4. Rotate the macro image by the difference (θ − φ) to achieve alignment.
5. Convert pixel distances in the macro image into the nanoindenter's stage coordinates for precise positioning.

Figures S6a–S6d illustrate this process, with the reference points highlighted.

# Z Alignment

To ensure accurate depth correlation based on optical focus, Z-alignment is performed as follows:

1. Identify a third reference point located at a different height on the sample surface.
2. Assume height variation is primarily due to sample tilt, and establish a linear relationship between the x, y positions and corresponding Z displacements.
3. Use this relationship to compute a transformation that compensates for height variation, enabling sharp focus at all intended indentation sites.

This alignment method is demonstrated in Figures S6a–S6d.

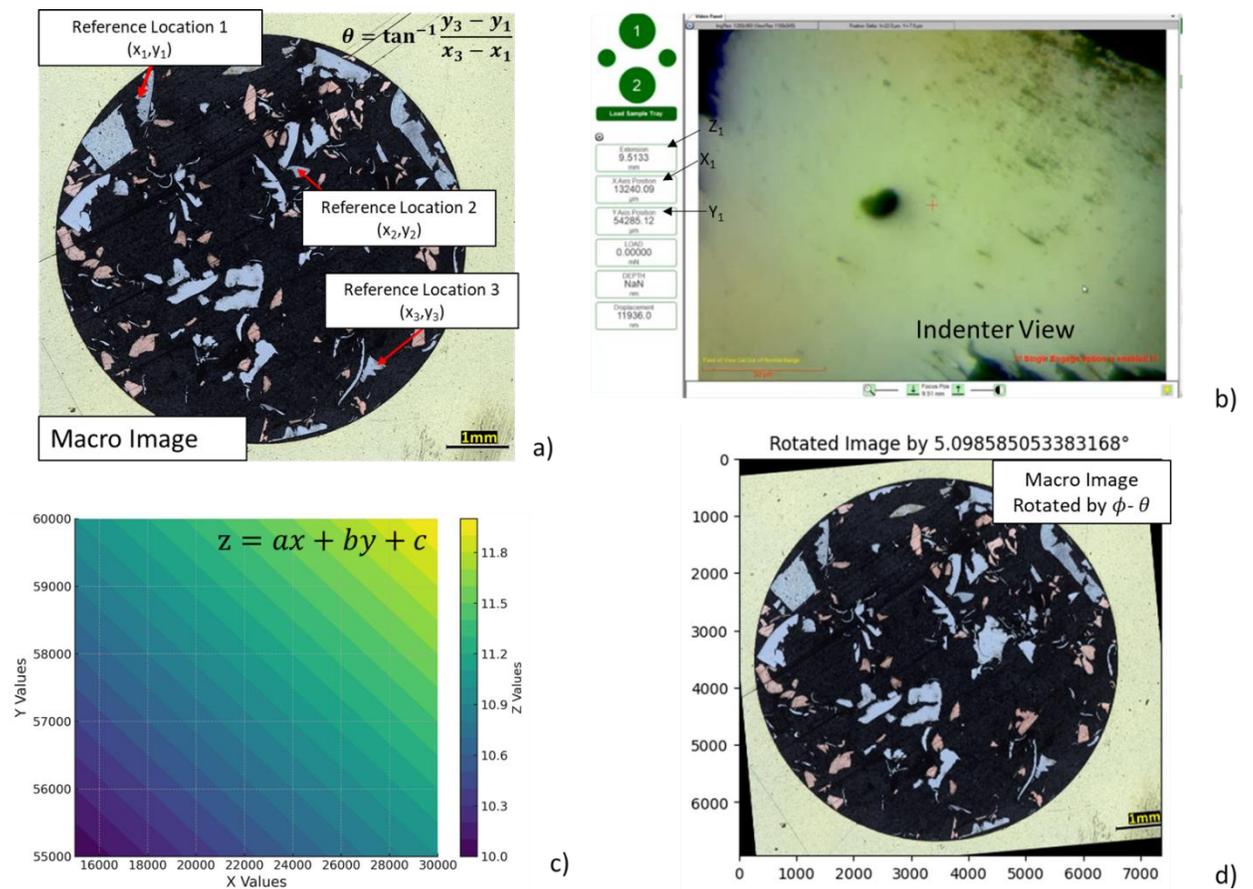

*Figure S6: a) Macro (Large) image from the optical microscope showing three reference locations and the angle between the two farthest features. (b) Indenter view at one of these locations with evaluation of the angle between the same features. (c) Z-position at the three locations used to determine the sample's focus through a linear relationship between x, y, and z. (d) Macro image rotated by the difference in angle identified in (a) and (b)*